# Enhanced Surface Plasmon Polariton Propagation Induced by Active Dielectrics

C. Athanasopoulos[1, 2], M. Mattheakis[*1, 2] and G. P. Tsironis[1, 2, 3]

[1] Crete Center for Quantum Complexity and Nanotechnology, Department of Physics, University of Crete, P. O. Box 2208, Heraklion 71003, Greece
[2] Institute of Electronic Structure and Laser, Foundation for Research and Technology-Hellas, N. Plastira 100, Vassilika Vouton, GR-70013, Heraklion, Greece
[3] Department of Physics, Nazarbayev University, 53 Kabanbay Batyr Ave., Astana 010000, Kazakhstan
*Corresponding author: mariosmat@physics.uoc.gr

**Abstract:** We present numerical simulations for the propagation of surface plasmon polaritons (SPPs) in a dielectric-metal-dielectric waveguide using COMSOL multiphysics software. We show that the use of an active dielectric with gain that compensates metal absorption losses enhances substantially plasmon propagation. Furthermore, the introduction of the active material induces, for a specific gain value, a root in the imaginary part of the propagation constant leading to infinite propagation of the surface plasmon. The computational approaches analyzed in this work can be used to define and tune the optimal conditions for surface plasmon polariton amplification and propagation.

**Keywords:** Surface Plasmon Polaritons (SPPs), Active dielectrics, Gain materials, Dispersion Relation, Propagation length.

## 1. Introduction

At optical frequencies the metal's free electrons can sustain, under certain conditions, oscillations, called surface plasmon polaritons (SPPs) or plasmons with distinct resonance frequencies [1-6]. The existence of plasmons is characteristic for the interaction of metals with light. Additionally many innovative concepts and applications of metal optics have been developed over the past few years [2-3][5][7]. Two commonly used configurations for plasmon excitation are the Kretschmann - Raether and the Otto configuration. In the first, a thin metal film (40-70nm) is sandwiched between two dielectrics, usually glass and air, with the incident wave hitting the denser medium. In the second, the denser dielectric and the metal sandwich the lighter medium [3-4][8].

The aim of this work is to investigate numerically, with COMSOL, the plasmon dispersion relation as well as propagation and the role of gain. We will use the Kretschmann - Raether configuration; the exact solution of Maxwell's equations can be taken from Ref. [1][3-6][9] for infinite planes in y-direction. However, a pure plasmon mode, as theory describes, can exist only theoretically due to boundary and initial conditions. That is to say, in experiments as well as in some COMSOL simulations, we have neither infinitive layers nor real plane wave source, resulting a mixed state of plasmons and photons propagating. In Fig. 1 we show a waveguide (layered structure) for plasmon excitation.

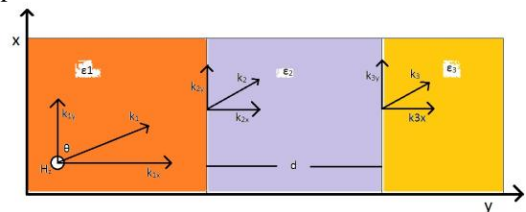

**Figure 1**. Configuration for surface plasmon excitation.

A metal film of thickness d and permittivity $\varepsilon_2$ is sandwiched between two dielectric layers with permittivity $\varepsilon_1$ and $\varepsilon_3$. The wave vector of the incident transverse magnetic (TM) wave is $k_1$ and the refracted wave vectors in the next two layers are $k_2$ and $k_3$. The permittivity of the metal is frequency dependent and is taken using the Drude-Sommerfeld theory (or Drude model) given by Eq. (1) [2-5][10], where $\varepsilon_h$ is the high frequency permittivity, $\omega_p$ is the plasma frequency, $\Gamma$ is the collision frequency and $\omega$ is the angular frequency of the EM wave.

$$\varepsilon_2(\omega) = \varepsilon_h - \frac{\omega_p^2}{\omega^2 - i\omega\Gamma} \qquad (1)$$

Note that the minus sign in term $i\Gamma\omega$ of Eq. 1 comes as consequence of the sign convention that COMSOL uses for time harmonic fields. In our analysis we use silver and two different types of silica glass; Table 1 shows the parameters used.

**Table 1:** Values of parameters

| Parameter | Value |
|---|---|
| $\varepsilon_1$ | 2.25 |
| $\varepsilon_3$ | 1.69 |
| D | 50 nm |
| $\Gamma$ | $1.018 \; 10^{14}$ rad/sec |
| $\omega_p$ | $1.367 \; 10^{16}$ rad/sec |
| $\varepsilon_h$ | 9.84 |

The COMSOL simulations based on [8] and theoretical calculations taken from [3-4] are presented in this section. The optimal incident angle corresponds at the minimum of the reflection response of the incident wave [3-4] [11-12]. Fig. 2 shows the theoretical prediction (solid line) of the reflection response as function of incident angle [3-4], whereas circles are COMSOL's results. At the resonance angle of $66.74^o$ the maximum amplitude surface plasmon is formed.

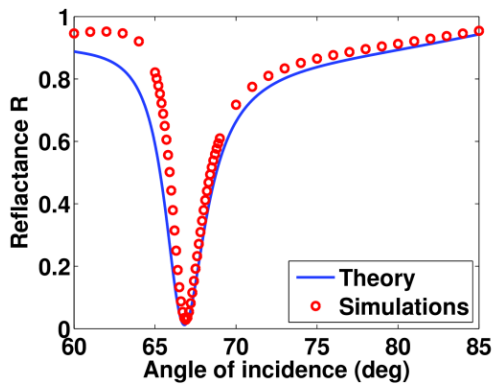

**Figure 2.** Angle dependence for the Kretschmann-Raether configuration: Analytical calculations (blue line) and COMSOL simulation (red circles). The resonance angle is at $66.74^o$ and corresponds to the angle for which the total reflection response is minimum.

In Fig. 3 we show the magnetic field distribution (xy plane) at resonance. Note that we have imposed Floquet periodic boundary conditions on both up and down boundaries for a better representation of the plasmon [8]. The major diagram is a color scaled image of the magnetic field while the top right insert is the corresponding 3D plot and the bottom right insert is the characteristic profile of the magnetic field distribution along the x-axis, across the materials [8].

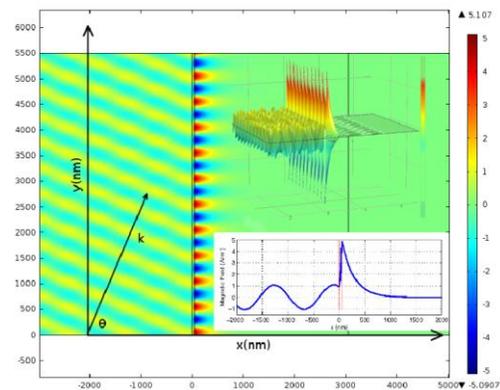

**Figure 3.** Magnetic field $H_z(x,y)$ at the resonance angle (COMSOL simulation). The major diagram is a color scaled image of the magnetic field while the top right insert is the corresponding 3D plot and the bottom right insert is the characteristic profile of the magnetic field distribution along the x-axis, across the materials

The plan of this paper is the follows: in the next section we investigate numerically the plasmon dispersion relation and we show that the numerical results agree with the theory. In section 3 we find again the dispersion relation of the plasmons with the presence of gain (active material) and we compare it with theoretical predictions. In section 4 we investigate the propagation length of SPPs and how the presence of an active material improves it. Finally, we conclude with basic results and potential fields of application.

## 2. Dispersion Relation

In this section we simulate the relation between the wave vector along the interface $\beta$

and the angular frequency ω; it is called dispersion relation and is given by Eq. 2:

$$\beta = k_y = \sqrt{\frac{\varepsilon_2 \varepsilon_3}{\varepsilon_2 + \varepsilon_3}} \frac{\omega}{c} \quad (2)$$

The normal component of the wavevector is

$$k_{j,x} = \sqrt{\frac{\varepsilon_j^2}{\varepsilon_2 + \varepsilon_3}} \frac{\omega}{c} \quad (3)$$

where j = 1, 2, 3 denotes the material [1-5][8].

In order to obtain an exponential decaying solution in x-direction, i.e. purely imaginary $k_x$, as well as a wave propagation solution in y-direction, i.e. real $k_y$, the following conditions must be satisfied [3-4]

$$\varepsilon_2(\omega) \cdot \varepsilon_3 < 0 \quad (4)$$

$$\varepsilon_2(\omega) + \varepsilon_3 < 0 \quad (5)$$

Metals, especially noble metals such as gold and silver, have a large negative real part of the dielectric constant along with a small imaginary part. Therefore, at the interface between a noble metal and a dielectric the conditions of Eq. 4 and Eq. 5 are satisfied and localized modes can exist. Fig. 4a and Fig. 4b show the dispersion relation (real and imaginary part respectively) for the materials. Points are COMSOL simulations, solid lines are theoretical plots of Eq. 2 while dashed lines are the photon's dispersion relation (ω = kc/n) in the two dielectrics with refractive index $n = \sqrt{\varepsilon_3}$ respectively.

## 3. Dispersion Relation with an Active Dielectric - Gain

We now introduce active materials, i.e. materials that have complex permittivities, where the imaginary part accounts for gain [10-11] [13-18]. Gain will counterbalance the ohmic loss of the metal in the SPP propagation. Table 2 shows the parameters used for our calculations.

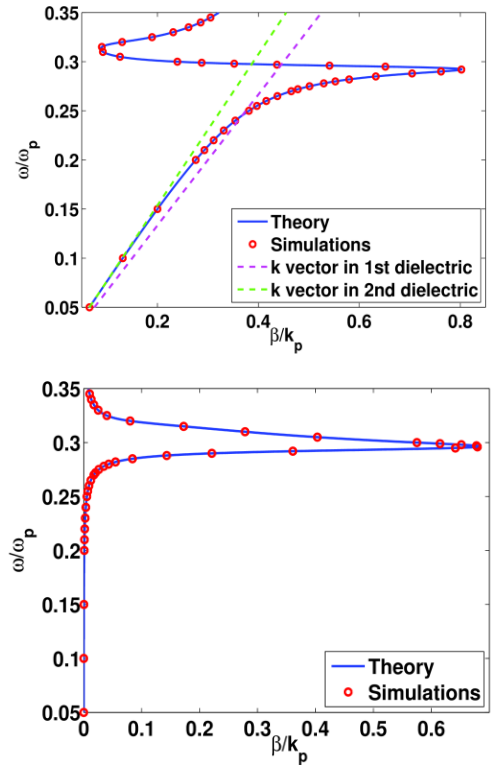

**Figure 4**. Dispersion relation. The top diagram (a) is the real part of the dispersion relation. The dashed purple line denotes the dispersion relation (k vector) in the first dielectric whereas the dashed green line is the dispersion relation for the second dielectric. The blue solid line is the real part of dispersion relation of SPPs based on Eq. 2 whereas the red circles show the COMSOL results. The latter diagram (b) indicated the imaginary part of SPPs dispersion relation based on Eq. 2. The blue solid line is the imaginary part of dispersion relation of SPPs based on Eq. 2 whereas the red circles show the COMSOL results. In both diagrams (a-b) $\omega_p$ and $k_p=\omega/c$ are normalization constants.

**Table 2:** Values of parameters (GAIN)

| Parameter | Value |
|---|---|
| ε1 | 2.25 |
| ε2 | -15.13 -0.93i |
| ε3 | (1.3+iκ)^2 |
| κ | Varies from 0 to 0.0099 |
| θ | 66.74^o |
| d | 50nm |

In the presence of an active dielectric with $\varepsilon_3 = (1.3 + i\kappa)^2$ the maximum amplitude of the z-component of the magnetic field is significantly greater as it can been seen in Fig. 5a, while the plasmon propagates longer in the y-direction, as can be seen in Fig. 5b [10-11][13][15][17].

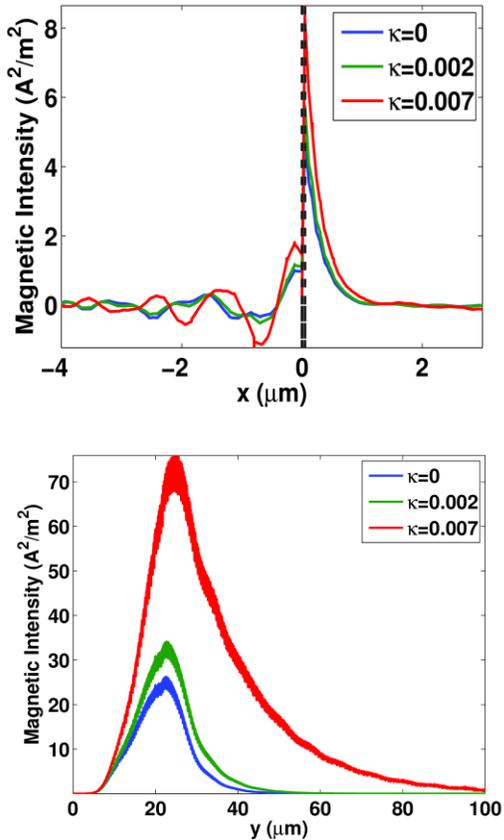

**Figure 5**. Intensity of magnetic field with and without the presence of gain. (a) Magnetic field across the materials for dielectric with no gain, κ=0 (blue line), same as in Fig. 3 and for gain κ=0.002 (green line) and κ=0.007 (red line). Dashed lines indicate the metal boundaries. (b) $H_z(y)$ on the interface for dielectric with no gain, κ=0 (blue line; bottom), and for gain κ=0.002 (green line; middle) and κ=0.007 (red line; top).

The real part as well as the imaginary part of the propagation constant β (dispersion relation), for complex relative permittivity $\varepsilon_3$, are shown in Fig. 6a and in Fig. 6b respectively. For comparison, the real and imaginary parts are shown when a purely real $\varepsilon_3$ is used. As a result, the introduction of an active dielectric shifts the curve of dispersion relation towards a greater wave vector β resulting longer propagation [10][13-14].

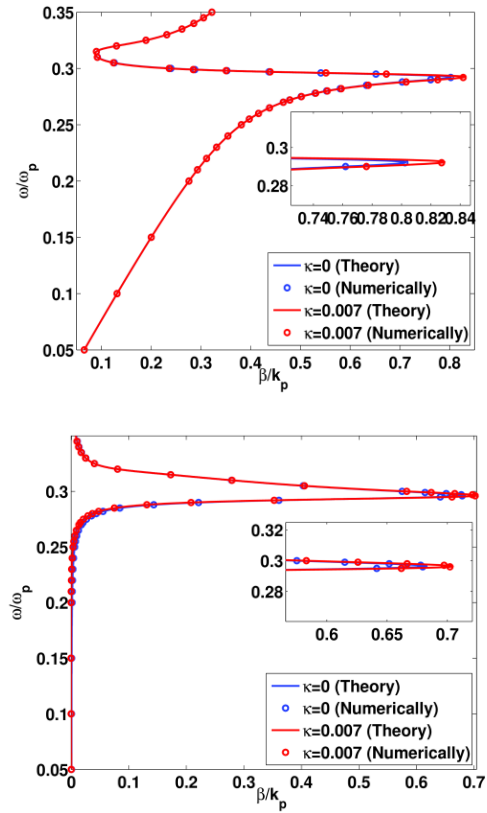

**Figure 6**. Dispersion relation with the presence of active dielectric. The red points are the COMSOL simulations while lines are analytical calculations based on Eq. 2. (a) Real part of β. Blue line: no gain; red line: gain κ = 0.007. (b) Imaginary part of β. Blue line: no gain; red line: gain κ = 0.007.

## 4. Plasmon Propagation along interface

In this section we simulate the plasmon propagation along the interface in the presence of active dielectrics with different gains κ [10-13][15][17]. If we consider the complex wavenumber $\beta = \beta' + i\beta''$, then the real part β determines the SPP wavelength, while the imaginary part β accounts for the damping of the SPP as it propagates along the interface.

The propagation lengths L and $L_F$ of the SPP along the interface corresponds to the length traveled by the plasmon up to 1/e decrease of the intensity or field amplitude and is determined by the imaginary part $\beta''$ from the Eqs. 6 and 7 for EM intensity and field respectively [2-3][5][14].

$$L = \frac{1}{2\,\text{Im}[\beta]} \quad (6)$$

$$L_F = \frac{1}{\text{Im}[\beta]} \quad (7)$$

Furthermore, we proceed with the measurements of the propagation length L for different values of gain κ to quantify the dependence of L as a function of the gain parameter κ.

In Fig. 7 we show the fits (straight lines on a semilog plot) of the intensity of the magnetic field along the interface for a few values of gain. The fits estimate the slopes that correspond to the propagation length. Results for various gains are presented in Table 3. As the imaginary part of the refraction index (gain) increases, the SPPs propagate longer along the interface. For example gain of 0.6% (imaginary part over real part of refraction index) facilitates plasmon propagation by a factor of three. At the limit when the imaginary part of the propagation constant tends to zero (ohmic losses are 'equal' to gain), plasmons propagate without losses and propagation length goes to infinity. The fit to an exponential curve is not appropriate. This critical gain, obtained numerically with COMSOL, gives a plasmon propagating at constant amplitude.

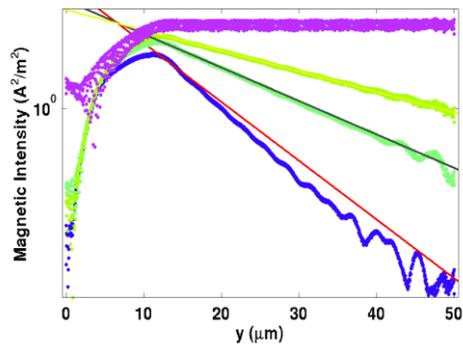

**Figure 7**. Fits on the logarithm of the intensity of the magnetic field component along one side (up) of the plasmon propagation. Red line: no gain; black line: gain κ = 0.005; yellow line: gain κ = 0.007. Top graph for gain κ = 0.0099 (no fit).

**Table 3:** Propagation length L as function of gain κ.

| Gain κ | Propagation Length L (μm) |
|---|---|
| 0 | 5.7 |
| 0.002 | 6.9 |
| 0.004 | 8.6 |
| 0.005 | 10.0 |
| 0.006 | 12.3 |
| 0.007 | 16.3 |
| 0.008 | 25.1 |
| 0.0085 | 34.2 |
| 0.009 | 53.9 |
| 0.0092 | 69.7 |
| 0.0094 | 99.3 |
| 0.0095 | 125.3 |
| 0.0096 | 169.5 |
| 0.0097 | 261.2 |
| 0.0098 | 564.7 |
| 0.0099 | ∞, constant amplitude |

The theoretical values for the propagation length derived from Eq. (6) or (7) differ from the ones obtained numerically by COMSOL. The reason, as mentioned earlier, is that the exact SPPs solution is derived for plane waves and semi-infinite SPP waveguides, while in the simulations we generate mixed plasmon mode. However, in both cases, i.e. theoretical and computational, functional dependence of the propagation length as a function of the gain is qualitatively similar. This relation is shown graphically in Fig. 8, where we can see that simulation results and theory give the same behavior, but disagree in quantification. We have also investigated the affect that the size of the source has on the propagation length. We used sources of various sizes (from 5 to 50 μm) and found that as the source size increases as the numerical results tends to theoretical predictions.

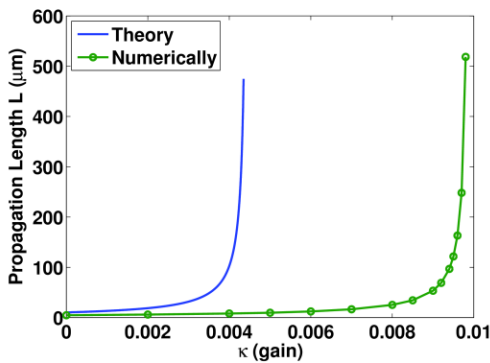

**Figure 8.** Plasmon propagation length for different gain values κ: theory (blue) and COMSOL simulations (green).

## 5. Conclusion

In this work we used COMSOL in order to simulate surface plasmon polaritons propagation (SPPs) in a dielectric-metal-dielectric waveguide (Kretschmann-Raether configuration). We added gain in the second dielectric in order to overcome metal losses. We introduced gain by adding a small imaginary part in the refractive index of the dielectric; values were of the order of $10^{-3}$ compared to its real part. We found that SPP propagation was enhanced along the SPPs transverse direction x-axis, but most importantly along the metal - dielectric interface, y-direction. We performed numerical fits and estimated the 1/e amplitude SPP propagation length to be 5.7μm in a non active dielectric and 53.9μm in an active dielectric with complex refractive index n=(1.3+0.009i), an order of magnitude larger. Specifically, we also found that for a given value of gain the propagation length becomes infinite. This property is due to a root in the imaginary part of SPP propagation constant β induced by the presence of gain. This feature was seen both analytically as well as numerically. The quantitative discrepancy between the theoretical and computational result is due to the specific SSP initial and boundary conditions used. The results of this work will be useful for designing plasmon devices with gain.

## 6. Acknowledgements

This work was supported in part by by the European Union grant FP7-REGPOT-2012-2013-1: CCQCN-316165. In addition it was partially supported by the THALES project "ANEMOS" co-financed by the European Union (European Social Fund-ESF) and Greek national funds through the Operational Program "Education and Lifelong Learning" of the National Strategic Reference Framework (NSRF) –Research Funding Program: THALES-Investing in knowledge society through the European Social Fund.